\documentclass[prb,twocolumn]{revtex4}% Physical Review B
\usepackage[T1]{fontenc}% pour utiliser les caractères français 
\usepackage{graphicx}% Include figure files 
\usepackage{dcolumn}% Align table columns on decimal point 
\usepackage{bm}% bold math 
\usepackage[dvips]{color}

\begin{document} 
\title{Density hardening plasticity and mechanical aging of silica glass
 under pressure: \\A Raman spectroscopic study}

\author{Damien Vandembroucq$^{1,3}$
\thanks{damienvdb@pmmh.espci.fr},
Thierry Deschamps$^2$, Camille Coussa$^2$, Antoine Perriot$^3$,
Etienne Barthel$^3$, Bernard Champagnon$^2$ and Christine
Martinet$^2$}

\affiliation{$^1$Laboratoire de Physique et Mécanique 
des Milieux Hétérogènes,\\
PMMH UMR 7636 CNRS/ESPCI/Paris 6/Paris 7,\\
10 rue Vauquelin, F-75231 Paris cedex 05, France.\\
\\
$^2$Université de Lyon; Université Lyon1; UMR5620 CNRS\\
Laboratoire de Physico-Chimie des Matériaux Luminescents, \\
Domaine Scientifique de la Doua, Bât. Kastler, \\10 rue Ampère, 
Villeurbanne, F-69622, France.\\
\\
$^3$Laboratoire Surface du Verre et Interfaces,\\
Unit\'e Mixte de Recherche CNRS/Saint-Gobain,\\
39 Quai Lucien Lefranc, F-93303 Aubervilliers, France.\\
}

\begin{abstract}

In addition of a flow, plastic deformation of structural glasses (in
particular amorphous silica) is characterized by a permanent densification.
Raman spectroscopic estimators are shown to give a full account of the
plastic behavior of silica under pressure. 
%Spectroscopic studies under
%high pressure thus help us to understand not only the amorphous
%structure but also the plastic behavior of silica. Raman scattering
%data can for instance be used to identify the plastic constitutive law
%of silica.  In parallel, the evolution of some patterns of the Raman
%spectrum in the course of a high pressure experiment can be analysed
%according to a mechanical point of view.  
While the permanent densification of silica has been widely discussed
in terms of amorphous-amorphous transition, from a plasticity point of
view, the evolution of the residual densification with the maximum
pressure of a pressure cycle can be discussed as a density hardening
phenomenon. In the framework of such a mechanical aging effect, we
propose that the glass structure could be labelled by the maximum
pressure experienced by the glass and that the saturation of
densification could be associated with the densest packing of tetrahedra
only linked by their vertices.

\end{abstract}
 
\date{\today}
\maketitle
% body of paper here

\section{Introduction}

%pour entrer une correction il suffit d'écrire \PCML{correction}
%\PCML{ceci est un exemple de correction}

%{\bf context 1 \--- effect of pressure: generalities} 
The behavior of amorphous silica under high pressure has been
intensively studied in the last
decades\cite{bridgman-JAP53,Cohen-JAP61,Mackenzie-JACS63,Polian-PRB90,Meade-PRL92,Inamura-JNCS01,Inamura-PRL04}
and has recently motivated an increasing amount of numerical
studies\cite{Kalia-PRL93,Kalia-PRB94,Venuti-PRB96,Lacks-PRL98,Lacks-PRL00,Benoit-PRB03,Garofalini-PRL03,Kieffer-PRB04a,Kieffer-PRB04b,Scandolo-PRB07}. Silica
appears to be elastic up to around 10 GPa and to exhibit a plastic
behavior at higher pressure. Two features can be emphasized at that
level: i) in the elastic regime, the compressibility exhibits a
surprising non monotonous evolution with a maximum at around 2-3
GPa\cite{bridgman-JAP53,Polian-PRB93,Zha-PRB94} ii) when unloading
from the plastic regime a permanent densification up to 20 \% can be
observed\cite{bridgman-JAP53,Cohen-JAP61,Mackenzie-JACS63,Polian-PRB90,Meade-PRL92,Inamura-JNCS01,Inamura-PRL04}. No
clear evidence of alteration of the tetrahedral short range order in
this unloaded state has been observed\cite{Mysen-PRB86,Meade-PRL92}.

%{\bf context 2 \--- Pressure induced structural transformations}
Above 25 GPa, a change from 4-fold to 6-fold coordination is
observed\cite{Meade-PRL92}. This 6-fold amorphous seems not to be
quenchable at zero pressure. When unloading downto zero pressure, no
trace of 6-fold coordination is obtained. Performing X-ray Raman
scattering experiments on oxygen K-edge, Lin {\it et al}
\cite{Lin-PRB07} observed a reversible electronic bonding transition
between 10 and 25 GPa. The latter was attributed to a fourfold quartz-like
to a sixfold stishovitelike change of configuration  of silica glass.
%\DV{Why are these results controversial ?}.  
 For $P>25 \;\mathrm{GPa}$ the densification process saturates and
after unloading to ambient, the density level is the one obtained with
a maximum pressure $P\simeq25$ GPa \cite{Polian-PRB93,Zha-PRB94}.

Questions remain about the nature of the densified phase and the
mechanism of densification. Recent studies\cite{Scandolo-PRB07} have
proposed the existence of an ``activated'' five-fold coordination at
high pressure allowing reorganization toward a denser tetrahedral
network. In former studies, in analogy with amorphous ice,
Lacks\cite{Lacks-PRL00} has proposed a first order transition between
2 different tetrahedral amorphous phases of silica. This transition
would be kinetically hindered at room temperature. This idea of
poly-amorphism has received a lot of attention
\cite{Sciortino-Nat01,Mukherjee-PRL01,Debenedetti-PRE02,Garofalini-PRL03,Sciortino-PRE04,Inamura-PRL04,Kieffer-PRB04a,Kieffer-PRB04b,Champagnon-JNCS07}. At
this stage it is important to separate the known transition at very
HP between a 4-fold amorphous silica and a 6-fold amorphous silica
involving a change in the short-range
order\cite{Meade-PRL92,Kalia-PRB94} from an additional hypothetical
transition at lower P between 2 different amorphous phases of
tetrahedral silica and involving the medium range
order\cite{Poe-JNCS04}.

%{\bf context 3 \--- Link with plasticity of amorphous materials} 

The most recent numerical\cite{Scandolo-PRB07} and
experimental\cite{Inamura-PRL04} works as well as the existence of a
continuous range of densities for amorphous
silica\cite{Inamura-JNCS01} after return to ambient pressure seem to
rule out this idea of transition between 2 different forms of
amorphous tetrahedral silica. However, the original observation of
Lacks remains of interest: performing molecular dynamics simulations
driven in volume\cite{Lacks-PRL98}, he noted discontinuities in the
pressure signal associated with local pressure induced mechanical
instabilities. The latter are reminiscent of the shear induced
mechanical instabilities previously identified in flowing
liquids\cite{Malandro-PRL98}. Note that similar localized transitions
are widely believed to be the main mechanism of shear plasticity of
amorphous materials\cite{FalkLanger-PRE98} and can be associated to
the vanishing of one eigenvalue of the Hessian matrix of the
interatomic potential\cite{Maloney-PRL04,Maloney-PRE06}.

%{\bf context 4 \--- Link with hardening and continuum plasticity } 

In parallel to a structural study, it may be worth considering the
pressure induced densification process in silica according to a
mechanical perspective. In particular, in the Raman spectroscopic
measurements to be presented below, we will use cycles of pressure of
increasing maxima. This protocol will help us to discriminate in the
spectral patterns, modifications due to the reversible elastic
deformation of the network from other ones due to plastic structural
reorganizations.

%{\bf Outline of the paper.}  

%%%%%%%%%%%%
We first present the experimental methods {\it i.e.}  Raman
measurements of silica submitted to pressure cycles and analyze their
results in terms of mechanical behavior. Two series of experiments are
discussed. In series {\bf A} Raman measurements are performed {\it in
situ} during the loading and unloading stages of successive pressure
cycles. In series {\bf B} Raman measurements are performed {\it ex
situ} at ambient pressure before and after pressure cycles of
increasing maximum pressure.

%%%%%%%%

The results of these spectroscopic measurements are presented in the
next section and discussed in the framework of continuum
plasticity. The first series of {\it in situ} experiments gives a nice
illustration of (densification) plasticity in the context of a silica
glass while the second series of experiments allows us to follow the
evolution of permanent densification {\it vs} the pressure maximum of
the cycle {\it i.e.} the density hardening behavior of silica.

%%%%%%%%%%

We finally give a discussion, first in terms of mechanical behavior,
then in terms of amorphous structure. These results are of primary
importance in the description of the mechanical properties of
silica. While silica is daily used as a calibration sample for
nano-indentation measurements, it appears that the mechanical behavior
of this material is not fully described. It has been shown
recently\cite{PMMCVB-JACS06} that the constitutive laws
available\cite{lamb2d,lamb3d} which do not take hardening into account
fail to fully describe the densification process induced by an
indentation test. The above data can be used to include hardening in a
simple constitutive law of silica\cite{KBVD-ActaMat08}, which gives a
precise account for this behavior.

Beyond their interest in terms of mechanical behavior, these results
can also be discussed in terms of amorphous structure. In particular,
as detailed below, densification can be regarded as a typical glassy
phenomenon, resulting from a mechanical aging process. Following this
perspective, the amorphous densified structure would be associated with
the quench of the structure at high pressure and could be labelled by
a fictive pressure in the very same acception as a fictive temperature
can be used to label a structure obtained by thermal aging.

\section{Experimental Methods}

Bars of amorphous silica (Saint-Gobain Quartz IDD) are shattered into
pieces. Splinters of characteristic length of 10 $\mu$m are submitted
to cycles of pressure in a ``Sidoine''\thanks{Laboratoire de Physique
des Milieux Condensés(Paris)} Diamond Anvil Cell (DAC) with a maximum
pressure in the range [1-25 GPa]. Raman spectra are collected with a
Renishaw RM 1000 micro-spectrometer with a Ar$^+$-ion 514 nm, 50 mW
laser excitation). A small piece of ruby is introduced together with
the silica splinter in order to monitor the pressure level using the
shift of the R1-luminescence band.

%\begin{figure}
%  \includegraphics[width=0.8\columnwidth]{sidoine.eps}\\
%\caption{Schematic representation of the used DAC apparatus. The
%sample and a piece of ruby were introduced into the compartment,
%previously caved into the steel joint using electroerosion. The ruby
%luminescence is used to measure the in-situ pressure applied by the
%diamonds. The presence a pressure transmitting fluid in the
%compartment ensures the quasi-hydrostaticity of the stress
%field.\PCML{Figure en anglais Membrane plutot que Chambre à
%air}}\label{sidoine}
%\end{figure}

Two series of experiments are presented. 
\begin{itemize}
\item 
In series {\bf A} {\it in situ} measurements of the Raman spectrum are
collected all along the compression cycle. Three cycles are presented,
The first one with $P_{max}=7.3 \;\mathrm{GPa}$ lies in the elastic
domain, te third cycle consists of a compression up to $P_{max}=18
\;\mathrm{GPa}$ followed by a decompression at 1 GPa, finally the last
cycle consists of a compression up to $P_{max}=16 \;\mathrm{GPa}$ and
a direct return to ambient pressure induced by the breakdown of a
diamond. Methanol is used as a pressure transmitting fluid in this
series of experiments (quasi-hydrostatic conditions).
%\DV{Add details about non-hydrostaticity of methanol at
%HP if needed}\PCML{On a utlisé le méthanol pur car le mélange
%éthanol-méthanol gène l'observation car des bandes de l'éthanol se
%superposent avec la silice}. 
A detailed presentation of these {\it in situ} experiments can be
found in \cite{Champagnon-JNCS08}.
\item
In series {\bf B} Raman measurements are performed{\it ex situ} at
ambient pressure before and after each cycle of pressure. A series of
compression cycles is presented where the pressure maximum is
increasing from 9 GPa for the first cycle to 25 GPa for the last
cycle. In this case, the pressure transmitting fluid is a mix similar
to 5:1 methanol-ethanol, which ensures hydrostaticity up to 16 to 20
GPa \cite{metheth}. The Raman spectrum is measured before loading and
after unloading, the diamond-anvil-cell being emptied of the
transmitting fluid.
\end{itemize}

\begin{figure}[b]
  \includegraphics[width=0.8\columnwidth]{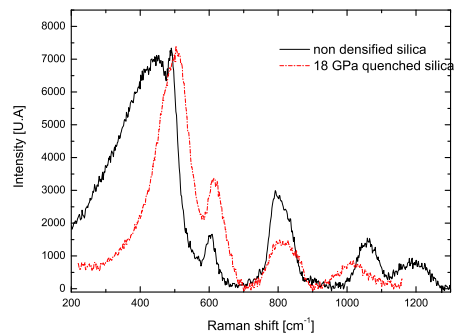}\\
\caption{Raman spectra obtained with an amorphous silica sample before
(plain line) and after (dotted line) a 18 GPa hydrostatic
loading.}\label{spectra}
\end{figure}

Figure \ref{spectra} contrasts the initial spectrum of a sample (plain
line) and that obtained after a 18 GPa hydrostatic loading (red line).

\textcolor{black}{Between 200 and 750 cm$^{-1}$ we can identify a main
band at 440 cm$^{-1}$. This band is intense, and affected by the
densification process: the band gets narrower and is shifted to higher
wave numbers. This band has originally been attributed to the
symmetric stretching mode of bridging oxygens between two Si atoms and
its evolution under densification to the decrease of the
inter-tetrahedral angles Si-O-Si \cite{Mochizuki-SSC72}. Recent
determinations of the Raman spectrum from first
principles\cite{Umari-PRL03} have established that this broad band
originates more likely from bending motions of oxygen atoms that do
not belong to small rings.}

The defect lines D$_1$ and D$_2$, at 492 and 605 cm$^{-1}$, are
respectively attributed to the breathing modes of the four-membered
and three-membered rings
\cite{Galeener-SSC82,Pasquarello-PRL98,Benoit-PRB03}. Their area ratio
was previously used in literature as an indicator for the variation of
the ``fictive temperature''
\cite{Champagnon-JNCS00,Champagnon-JNCS01,Champagnon-PMB02}, which is
associated with a slight change of density. The effect of pressure
seems to be better accounted for by the shift of the D$_2$ line. As
discussed by Polsky {\it et al}\cite{Polsky-JNCS99}, this may be due
to pressure induced variations of the Raman cross section.  The D$_2$
line is of particular interest since it has almost no overlap with the
main band. Sugiura et al. \cite{Sugiura-JAP97} correlated the position
of the D$_2$ line with the ratio of the sample density $\rho$ to its
initial density $\rho_0$. The residual density evaluated through this
relation accounts for both irreversible and elastic densification due
to residual elastic strain. However, several works
\cite{Hibino-APL85,Michalske-PCG88,Sugiura-JAP97} evidenced that the
D$_2$ line position is only marginally sensitive to residual elastic
strains.  This correlation was recently used to probe the
densification gradient surrounding a plastic imprint in silica
obtained by indentation\cite{PMMCVB-JACS06}.

\begin{figure}[b]
  \includegraphics[width=0.8\columnwidth]{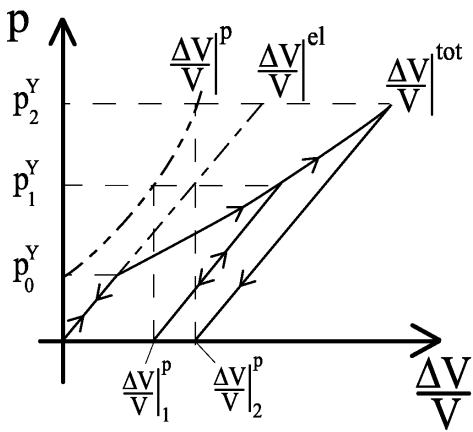}\\
\caption{Sketch of a typical pressure/density curve expected in
elastoplasticity. A reversible elastic behavior is obtained up to the
elastic limit pressure $P^Y_0$. When loading above $P^Y_0$, plasticity
sets in and the (elastic) unloading from $P^Y_1$ is characterized by a
residual densification. A subsequent loading at $P^Y_2>P^Y_1$
reproduces the previous unloading curve up to $P^Y_1$ before plasticity
sets in again. The elastic limit has thus evolved under loading from
$P^Y_0$ to $P^Y_1$ and $P^Y_2$. The knowledge of this density
hardening behavior (evolution of the limit elastic pressure with
density at zero pressure) is necessary to give a proper modelling of
plasticity of glasses.\label{elastoplast-curve}}
\end{figure}

\section{Results and interpretation}

We describe in the following the experimental results and give an
interpretation in the framework of the elasto-plastic response of
continuous media.

%\begin{figure}[h!]
%  \includegraphics[width=0.7\columnwidth]{hardening.eps}\\
%\caption{Limit of the elastic domain in the space of stresses
%(pressure in abscissa, equivalent shear stress in ordinate). The
%behavior remains elastic for any stress left or below this limit. When
%stress reaches this limit, plasticity sets in and the limit curve can
%be convected at higher stresses (hardening). Pressure cycles allow to
%give a quantitative account of the evolution of the limit pressure $P^Y$
%with density.\label{hardening}}
%\end{figure}

\begin{figure}[b]
\begin{center}
%\includegraphics[width=0.8\columnwidth]{cycle-BP.eps} 
%\vspace{1cm}
\includegraphics[width=0.95\columnwidth]{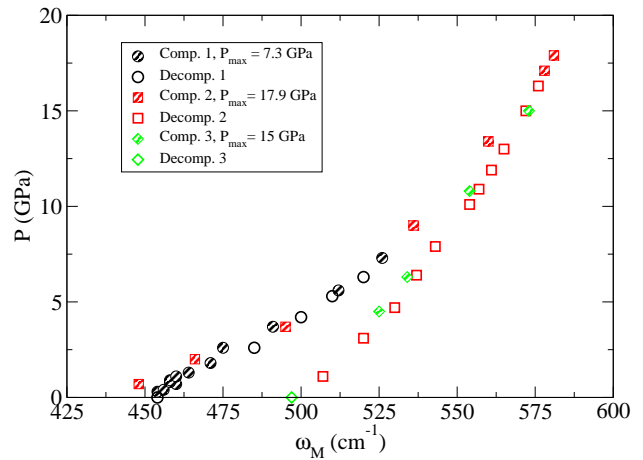} 
\end{center}
\caption{\label{Raman-cycle-BP} Raman shift of the main band
$\omega_M$ in the elastic-plastic regime. Successive pressure loading
and unloading cycles are depicted. Filled and empty symbols correspond
to loading and unloading curves respectively. For a  pressure maximum
$P_{max}=7.3$ GPa, a full reversibility is obtained (elastic
behavior). For a larger maximum pressure $P_{max}=18$ GPa, a residual
shift is obtained after unloading (plastic behavior). An additional
loading at $P_{max}=16$ GPa reproduces the last unloading curve,
indicating the increase of the elastic limit under loading
(hardening). }
\end{figure}

\subsection{Continuum mechanics}
We first recall briefly the formalism of
elasto-plasticity. Fig. \ref{elastoplast-curve} shows the hydrostatic
stress as a function of the volumetric strain of a medium submitted to
pressure. Below a threshold value of pressure $P_0^Y$ the material
remains fully elastic and the volumetric deformation is
reversible. When increasing the pressure beyond threshold, plasticity
sets in and an irreversible deformation adds up to the elastic
reversible deformation. When unloading from $P_1^Y>P_0^Y$, the
material behaves elastically and only the elastic part of deformation
is recovererd. Loading again to a higher pressure $P_2^Y>P_1^Y$, we
observe the same phenomenology with a crucial difference: the onset of
plasticity has increased from $P_0^Y$ to $P_1^Y$. In other words, the
mechanical behavior depends on the history of mechanical loading. The
material has experienced hardening\cite{Khan-book95} which can be
regarded as a mechanical aging. Such a behavior is standard for metal
shear plasticity and usually results at the structural level from the
entanglement or the pinning of dislocations by
impurities\cite{Cottrell-book56}. Metals however do not exhibit any
volumetric plastic deformation (dislocation motion is a volume
conserving mechanism). However, irreversible changes of density are
familiar in soil mechanics and granular materials (dilatancy
effect)\cite{Jaeger-RMP96}.

The typical mechanical behavior of a material like silica glass can be
summarized as follows. Before any loading, the material is elastic up
to a threshold which depends both on pressure and shear. Using
hydrostatic pressure $p$ and equivalent shear stress $\tau$ as
coordinates, this elastic threshold corresponds to a continuous curve
intersecting the two axis.
When reaching threshold, if the stress is
increased, plasticity sets in and the elastic limit is convected to
the maximum value of stress experienced by the material. At
macroscopic scale, one tries to characterize this hardening behavior,
relating the evolution of densification with the  pressure maximum.
% aswell as (for instance) the plastic shear strain with the maximum shear
%stress\PCML{pas très clair}.  
At microscopic scale, in the present case of silica, in absence of a
microscopic mechanism as well defined as the motion of dislocation, a
pending question remains to identify and understand a structural
signature of density hardening.

%\vspace{0.5cm}
\subsection{Elasto-plastic behavior of silica under pressure}

In this section we present the results of {\it in situ} raman measurements (series {\bf A}). Two patterns of the Raman spectra measured {\it in situ}
during the pressure cycles are discussed: {\it i)} the shift of the
main band $\omega_M$; {\it i)} the shift of the $D_2$ line
$\omega_{D2}$. 

\begin{figure}[t]
\begin{center}
%\includegraphics[width=0.8\columnwidth]{cycle-D2.eps}
%\vspace{1cm}
\includegraphics[width=0.95\columnwidth]{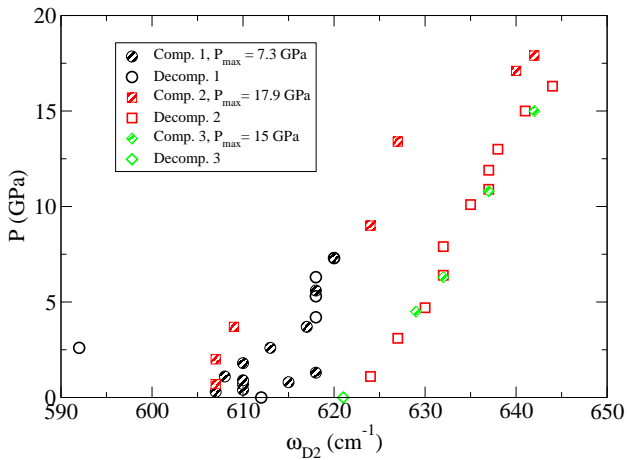} 
\end{center}
\caption{\label{Raman-cycle-D2} Raman shift of D2 line $\omega_{D2}$
in the elastic-plastic regime (same conditions as above).}
\end{figure}

On Fig. \ref{Raman-cycle-BP} and \ref{Raman-cycle-D2} we show the evolution of
the position of these two bands for a cycle of pressure up to
$P_{max}=7.3 \;\mathrm{GPa}$ (black symbols). We observe a full
reversibility between loading and unloading. This result is consistent
with the usual estimate $P_c\simeq 10 \;\mathrm{GPa}$ for the onset of
permanent densification in silica. The evolution of these indicators is also
plotted on the same figures for the second series of cycles up to
$P_{max}=17.9 \;\mathrm{GPa}$. We now see that below $P_c\simeq 10
\;\mathrm{GPa}$, the same elastic behavior as before is recovered; then a
change of slope on loading is noticeable at least for the position of
the main band; then the unloading curve does not reproduce the initial
loading one and there appears a permanent change of the
spectrum at ambient pressure; when loading again, one follows the very
last unloading curve. These two curves
appear to be very similar to the ideal case of hardening plasticity
depicted in Fig. \ref{elastoplast-curve}. 
%This is especially the case
%for the evolution of the D$_2$ shift which is known to depend
%monotonously on density. The relationship between density and the
%location of the main band is less clear. Indeed, densification is
%expected to be associated with changes of the medium range order, in
%particular in the distribution of tetrahedra rings[Ref]. Within these
%limitations, 
This spectroscopic study allows us to closely follow the
elastic-plastic behavior of silica under pressure.

%\begin{figure}
%\begin{center}
%\includegraphics[width=0.8\columnwidth]{elastic-silica-BP.eps} 
%\end{center}
%\caption{\label{Raman-elastic-BP} Raman shift of BP in the elastic regime}
%\end{figure}

%\vspace{0.25cm}

%\begin{figure}
%\begin{center}
%\includegraphics[width=0.8\columnwidth]{elastic-silica-D2.eps}
%\end{center}
%\caption{\label{Raman-elastic-D2} Raman shift of D2 in the elastic regime}
%\end{figure}

%\subsection{Density hardening}

\begin{figure}[t]
\begin{center}
\end{center}
%\vspace{1cm}
\begin{center}
\includegraphics[width=0.95\columnwidth]{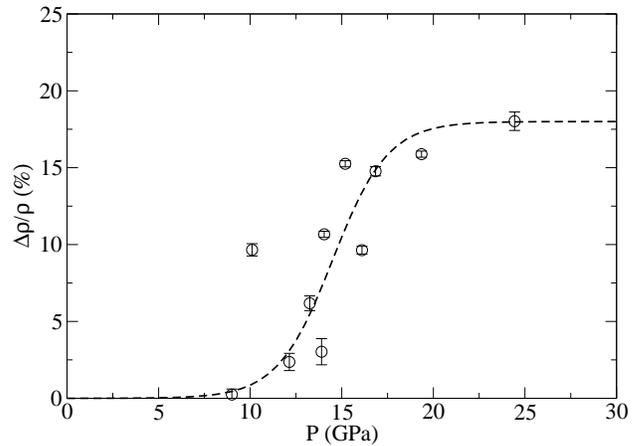}
\end{center}
\caption{\label{DAC-hardening} Evolution of the residual densification
vs the maximum pressure of the pressure cycle (symbols). The dashed
line is an indicative sigmoidal curve corresponding to a 18\% maximum
densification.}
\end{figure}

We now discuss the evolution of permanent densification after cycles
of increasing pressure. \textcolor{black}{Even if the signal-to-noise
ratio is less favorable for the determination of the D$_2$ line than
for the main band (see Fig. \ref{Raman-cycle-BP} {\it vs}
Fig. \ref{Raman-cycle-D2}), we choose the former to estimate the
permanent densification. As discussed above, the main reason for this
choice is that in contrast to the main band, the shift of the D$_2$
line appears to be rather independent of the elastic stress. This
allows us to use it a density probe in samples affected by residual
elastic stress. Because of the microscopic size of the silica samples
used in the diamond anvil cell, a quantitative calibration could not
be performed. Densification can be obtained on millimetric samples but
require high temperature treatments\cite{Inamura-JNCS01} and it is far
from being obvious that the medium range order is directly comparable
to the one obtained under high pressure at ambient temperature.}
Following Ref. \cite{PMMCVB-JACS06}, the densification was estimated
using the empirical relation:
\begin{equation}
\frac{\Delta \omega_{D2}}{\omega_{D2}} 
\simeq \left(\frac{\Delta \rho}{\rho} \right)^{0.14} \;,
\end{equation}
\textcolor{black}{which was extracted from the experimental data of
Sugiura et al. \cite{Sugiura-JAP97} obtained on shock wave
experiments. This calibration step is thus only approximative. }

The results are summarized on Fig. \ref{DAC-hardening}. We thus obtain
the evolution of the permanent densification with the pressure maximum
$P^Y$ of each pressure cycle. We obtain a continuous range of
increasing densities with an apparent saturation. As shown on
Fig. \ref{DAC-hardening}, this evolution can be approached by a
sigmoidal curve. Comparable results were obtained recently on window
glass using an octahedral multi-anvil
apparatus\cite{Rouxel-ScriptaMat06}. In mechanical terms, this allows
us to give a quantitative account of the density hardening behavior of
silica. In the space of stresses, we
have obtained the evolution of the maximum pressure $P_Y$ below which
the material remains fully elastic as a function of the glass density.

\section{Discussion}

Raman spectroscopic estimators have been shownn to give a full account
of the density hardening behavior of silica under pressure. Raman
scattering measurements during a loading/unloading pressure cycle
closely reproduce the elastoplastic behavior usually observed in
stress/strain curves: full reversibility below a limit stress,
appearance of a residual deformation with an elastic unloading for
larger stresses. A crucial difference obviously lies in the fact that
in the present study the evolution of hydrostatic pressure {\it vs}
density is considered instead of shear stress {\it vs} shear strain as
in usual metal plasticity.  From a mechanical point of view, the
detailed analysis of the density increase with the maximum pressure of
the cycle allowed us to study the evolution of the residual
densification with the elastic limit pressure $P^Y$. Such a knowledge
is of crucial importance for the determination of constitutive
equation modelling the plastic behavior of silica. It was shown in
Ref. \cite{PMMCVB-JACS06} that when restricting the plastic criterion
to perfect plasticity ({\it i.e.} assuming no hardening effect) it was
not possible to account for the permanent densification of amorphous
silica around a plastic imprint induced by indentation. Conversely, as
shown in Ref. \cite{KBVD-ActaMat08} the assumption of an elliptic
plastic criterion coupling shear stress and pressure together with the
data of density hardening extracted from the present experiments allow
to describe successfully this phenomenon of indentation induced
densification.

From a physical/structural point of view, the present results suggest
that the description of densified silica need not the hypothesis of an
amorphous-amorphous transition between two types of tetrahedral
networks. A simpler and alternative scenario consists of pressure
induced reorganizations of the amorphous network allowing a more
efficient packing of tetrahedra remaining linked by their vertices
only.  Such a scenario does not exclude the occurence of 5-fold or
6-fold coordinated silica in the plastic regime at high
pressure. However, the latter would correspond to intermediate states
between two amorphous tetrahedral structures. This occurrence of
5-fold or 6-fold coordination would thus simply denote the necessity
of cutting and rebonding between the two structures.  The denser
structure would thus be quenched when pressure decreases down to
ambient conditions. In that sense, the final structure could be
labelled by the maximum pressure the material experienced. In this
context of mechanical aging, the latter pressure could be thought of
as a ``fictive'' pressure in the same acception as the fictive
temperature in a more classical thermal aging experiment. Note that
these denser structures may be affected by internal stresses due to
the succession of localized reorganizations. \textcolor{black}{More
generally, these results indicate that the structure and the density
of a densified sample of vitreous silica will depend crucially on the
particular path it has followed in the plane pressure/temperature : it
is likely that the medium range sructure of densified silica is not
fully characterized by the only density parameter.}

Looking finally at
orders of magnitude for the density of the various phases of silica,
we observe that the density of the stable crystalline equivalent at
zero pressure (quartz) is 2.65 while the density of the metastable
coesite (stable at 2-3 GPa) is 3.01. The density of fused silica being
2.2, a 20\% increase gives 2.64 i.e values close to quartz but well
below coesite. In analogy with hexagonal and random close packing, an
interesting question could thus be whether the maximum observed
density of amorphous silica corresponds to any geometrical maximal
packing of tetrahedra bonded by their vertices only {\it i.e.} the
maximum density of the continuous random network. To our knowledge,
though the packing of space by tetrahedra, or ellipsoids has been
considered\cite{Kob-book05,Chaikin-PRL04}, this question has not been
discussed yet.

%\section{FEM Implementation, constitutive equation}

%\begin{figure}
%    \begin{center}
%      \includegraphics[width=0.8\columnwidth]{critere_silice.ps}
%      \end{center} \caption{The proposed yield criterion for amorphous
%      silica} \label{fig-critere} 
%\end{figure}

%\section{Conclusion}

\vspace{0.5cm}

\section{Acknowledgments}

The authors acknowledge useful discussions with G. Kermouche, W. Kob
and S. Ispas-Crouzet. They would like to thank CECOMO and High
Pressure facilities of University Lyon1. This work has been supported
by the ANR "plastiglass" contract n$^o$ ANR-05-BLAN-0367-01.

\bibliography{vdb,plasticity,silica,AP}

\end{document}